# IRCS/Subaru Observations of Water in the Inner Coma of Comet 73P-B/Schwassmann-Wachmann 3:

## Spatially Resolved Rotational Temperatures and Ortho-Para Ratios

Boncho P. Bonev [a, b, *], Michael J. Mumma [a], Hideyo Kawakita [c], Hitomi Kobayashi [c], Geronimo L. Villanueva [a, d]

**ACCEPTED FOR *ICARUS***

[a] *Solar System Exploration Division, NASA's Goddard Space Flight Center, Greenbelt, MD 20771, U.S.A.*

[b] *Department of Physics, The Catholic University of America, Washington DC 20064, U.S.A.*

[c] *Department of Physics, Faculty of Science, Kyoto Sangyo University, Motoyama, Kamigamo, Kita-ku, Kyoto 603-8555, Japan*

[d] *NASA Postdoctoral Fellow*

* *Corresponding Author. E-mail Address: Boncho.P.Bonev@nasa.gov*

"Water in the Inner Coma of 73P-B/Schwassmann-Wachmann 3"

Boncho P. Bonev

Building 2, Room 168, Code 693

NASA's Goddard Space Flight Center

Greenbelt, MD 20771

Email: Boncho.P.Bonev@nasa.gov

Phone: 301-286-1804

Fax: 301-286-0212

**Abstract**

Comet 73P-B/Schwassmann-Wachmann 3 was observed with IRCS/Subaru at geocentric distance of 0.074 AU on UT 10 May 2006. Multiple $H_2O$ emission lines were detected in non-resonant fluorescence near 2.9 μm. No significant variation in total $H_2O$ production rate was found during the (3 hour) duration of our observations. $H_2O$ rotational temperatures and ortho-to-para abundance ratios were measured for several positions in the coma. The temperatures extracted from two different time intervals show very similar spatial distributions. For both, the rotational temperature decreased from ~110 K to ~90 K as the projected distance from the nucleus increased from ~5 to ~30 km. We see no evidence for OPR change in the coma. The $H_2O$ ortho-para ratio is consistent with the statistical equilibrium value (3.0) for all spatially resolved measurements. This implies a nuclear spin temperature higher than ~ 45 K.

1.   **Introduction**

Water presents a paradigm for the release of parent[1] volatiles in comets. As the dominant volatile species in the nuclei of active comets, sublimation of water controls the release of other molecules within ~3 AU from the Sun. Understanding the excitation of $H_2O$ and other molecules in cometary comae is absolutely essential for determining accurate abundances of sublimated parent species, thereby characterizing the volatile composition of the comet nucleus.

Airborne and space telescopes have targeted primarily emissions in the water infrared (IR) fundamental bands (e.g., Mumma et al. 1986, Crovisier et al. 1997) or pure rotational transitions (e.g. Lecacheux et al. 2003, Bensch & Bergin 2004). The $H_2O$ molecule is now commonly observed from ground-based facilities via non-resonant ("hot-band") fluorescence in the near-IR[2] (e.g. Mumma et al. 1995, 1996; Dello Russo et al. 2004, 2005; Kawakita et al. 2006; Bonev et al. 2007).

Here we present results obtained from a detailed spectroscopic analysis of $H_2O$ "hot-band" emission from the inner coma of comet 73P-B/Schwassmann-Wachmann 3 (hereafter, SW3-B)[3]. The principal goal of this work is to investigate the (possible)

---

[1] The term "parent" (or "native") relates to molecules that have sublimated directly from the comet nucleus. Some molecules present as ice in the nucleus can also be released from extended sources (e.g. icy or dust grains) in the comet coma (escaping atmosphere). Chemical "splitting" (photodissociation, photoionization, etc) of parent molecules in the coma gives rise to "daughter" species (e.g. OH from $H_2O$).

[2] First detected in the Kuiper Airborne Observatory spectra of 1P/Halley (Mumma et al. 1986, Weaver et al. 1987).

[3] The Jupiter Family comet 73P/Schwassmann-Wachmann 3 (orbital period of ~5.4 yr.) split into three components ("A", "B", and "C") during its 1995 apparition and more components were found later (Sekanina 2005). The 2006 apparition was especially favorable (closest approach to Earth within < 0.1 AU). Component B was particularly

variation of $H_2O$ rotational temperature ($T_{rot}$) in the inner coma and the invariance of the ortho-para abundance ratio (OPR) after release from the nucleus. We do this by comparing spatially resolved measurements of the $H_2O$ rotational temperature and OPR. We also measure the $H_2O$ production rate ($Q(H_2O)$) for three separate time frames in search of temporal variability.

The rotational temperature probes conditions in the inner cometary coma. The OPR measurement relates to an important problem in physics (understanding the conditions for nuclear spin conversion; see Limbach et al 2006, Chapowski & Hermans 1999), and has also been considered to preserve information for the chemical formation temperature of cometary $H_2O$ (Crovisier 1984; Mumma et al. 1986, 1987, 1993). This (somewhat controversial) hypothesis implies that the OPR might have cosmogonic significance, since comet nuclei preserve a chemical record from the formative stages of our planetary system.

Section 2 gives the essential physics background (including defining terminology) for the excitation of water in cometary comae and the "hot-band" fluorescence mechanism. Section 3 describes our observations. Section 4 presents our detections of $H_2O$ in non-resonant fluorescence. Section 5 describes the spatially resolved measurements of $H_2O$ rotational temperature and OPR. Section 6 presents time-resolved $H_2O$ production rates, followed by a summary of results.

---

active in 2006 displaying frequent outbursts and fragmentations (see Weaver et al. 2006, Fuse et al. 2007). Various components of SW3 have been investigated at multiple wavelengths. Near-IR (2-5 μm) molecular studies focused primarily on the organic parent volatile composition of fragments "B" and "C" (Villanueva et al. 2006, Dello Russo et al. 2007, DiSanti et al. 2007, Kobayashi et al. 2007).

2. **Physics Background**

When a comet nucleus enters the inner solar system, sublimation processes are activated by sunlight. Sublimated molecules leave the nucleus with vibrational and rotational temperatures expected to be equal to the nucleus surface temperature, i.e. about 150-200 K at 1 AU. Except within a few hundred meters from the nucleus, collisional rates are too small to maintain the initial vibrational population, so this population decays radiatively to the ground vibrational state (Weaver & Mumma 1984). At 1 AU and for moderately high gas productivity (~$10^{28} - 10^{29}$ molecules sec$^{-1}$) rotational levels (in the ground vibrational state) are thermalized for distances up to a few thousand km, which usually exceeds the spatial extent of the coma sampled by near-IR observations (see Bockelée-Morvan & Crovisier 1987, Xie & Mumma 1992, and Bensch & Bergin 2004 for additional details).

Vibrational excitation of $H_2O$ molecules in the coma occurs mainly through infrared pumping by incident solar radiation. Non-resonant fluorescence embodies subsequent radiative decay into intermediate (between the ground and the pumped state) vibrational levels giving rise to the observed "hot-bands". These intermediate vibrational states are not significantly populated in the Earth's atmosphere, permitting $H_2O$ non-resonant fluorescence to be observed from ground-based observatories (see Mumma et al. 1995, 1996; Dello Russo et al. 2004 and references therein). Several hot-bands are now used, the most fruitful regions being near 2.9 and 4.7 μm.

Three physical parameters are derived from the analysis of $H_2O$ non-resonant fluorescence spectra: rotational temperature, ortho-para ratio, and $H_2O$ production rate. $Q(H_2O)$ [molecules s$^{-1}$] describes the total rate of release of water molecules from the comet nucleus. $T_{rot}$ is defined as the Boltzmann temperature that best reproduces the rotational energy distribution within the ground vibrational state of $H_2O$. Measurements of the near-nucleus temperature distribution are essential for understanding the excitation conditions in the inner collisionally-dominated cometary coma.

The $H_2O$ molecule is organized into two distinct spin species depending on whether the nuclear spin vectors of its (identical) hydrogen atoms are parallel (ortho "ladder") or anti-parallel (para "ladder"). The ratio between the total populations of ortho and para states (i.e. the OPR) is temperature dependent if we assume that the given OPR was realized in thermal equilibrium. Thus, nuclear spin temperatures ($T_{spin}$) can be derived from measured OPRs. Para states are increasingly favored at low spin temperatures, while at $T_{spin}$ > ~50 K the OPR reaches the statistical equilibrium value of 3.0 (for additional details see Mumma et al. 1987, Figure 4; Crovisier 2006; Bonev et al. 2007, Figure 1).

An important problem in chemical physics and in cometary science is whether or not water molecules undergo nuclear spin conversion during their long residence in the interior of a comet nucleus or during and after sublimation. If realized, such processes would alter the "primordial" OPR, leading to spin temperatures that differ from the chemical formation temperature of cometary $H_2O$. For an isolated $H_2O$ molecule in the coma, nuclear spin conversion is strongly forbidden for either radiative or (neutral-neutral) collisional transitions, leading to the prediction that the ortho-to-para ratio should be invariant with distance from the nucleus. Bonev et al. 2007 recently applied a method for testing nuclear spin conversion in the inner cometary coma, through spatially-resolved OPR measurements about the nucleus (the same technique is used in the present work).

3. **Detection of H$_2$O in Non-Resonant Fluorescence**

Infrared H$_2$O emission from the coma of comet SW3-B was observed on UT 10 May 2006 with the 8-meter Subaru telescope atop Mauna Kea, Hawaii. Spectra of high resolution ($\lambda/\delta\lambda \sim$ 10-12,000) were acquired with the Infrared Camera and Spectrograph (IRCS; Kobayashi et al. 2000). Based on these data, Kobayashi et al. (2007; hereafter Paper I) measured the organic volatile composition of SW3-B. In this second paper, we present a detailed spectroscopic study of H$_2$O "hot-band" emission. Table 1 presents a condensed observing log (see Paper I for detailed description). H$_2$O emission was sampled in three instrument settings (LA$^{+'}$, LA$^0$, and LA$^-$) defined by specific echelle/cross-disperser configurations.

<TABLE 1>

Two factors especially establish the value of the IRCS H$_2$O spectra. First, the data on UT 10 May 2006 offer broad spectral coverage within a single night. Quantum levels with a broad range of rotational energies (especially in LA$^-$) are sampled, allowing accurate retrieval of rotational temperature. Second, the comet was observed at geocentric distance of only 0.074 AU. This favorable observing geometry provided a rare opportunity to probe excitation conditions in the coma by "zooming" very close (~30 km) to the comet nucleus.

Algorithms for data processing, flux calibration (based on observation of standard star, see Paper I), and spectral extraction were developed specifically for our comet observations. These algorithms, along with recent substantial improvements are described elsewhere (Bonev 2005 (Appendix 2), Villanueva et al. 2006, DiSanti et al.

2006, and references therein). Figures 1a-1c show detections of multiple $H_2O$ lines sampled in the $LA^{+'}$, $LA^-$, and $LA^0$ setting respectively. The spectra are presented in a three-panel graphic. The top panels show the measured cometary spectrum and a best-fit atmospheric transmittance model normalized to the cometary dust continuum. The middle panels show residual spectra after removing the telluric component, revealing cometary gas emission alone. The bottom panels show modeled $H_2O$ spectra and will be discussed in §5.

<FIGURE 1a, 1b, 1c>

Telluric absorption lines appear against the comet continuum, and a very critical step in the spectral analysis is the proper synthetic modeling of the terrestrial atmospheric component. The importance of this modeling is two-fold. First, a proper continuum subtraction is essential for obtaining accurate residual spectra. Second (and equally significant), the fully resolved atmospheric transmittance function is later needed to correct the measured line fluxes for telluric extinction at their exact Doppler-shifted frequencies. In 2006 (Villanueva et al. 2006) we instituted the GENLN2 ver.4 line-by-line code (Edwards 1992) for synthesizing telluric transmittance, and we now use the HITRAN-2004 molecular database with 2005 updates (Rothman et al. 2005).

4. **Rotational Temperature and Ortho-Para Ratio**

4.1 *Methodology*

The rotational temperature and OPR can be determined from the relative intensities of the detected $H_2O$ lines (see Bonev et al. 2007, DiSanti et al. 2006, Dello

Russo et al. 2004). Some ortho and para transitions that are closely spaced in frequency appear blended at the resolution of our spectra. In order to use all $H_2O$ lines (including ortho-para blends) in the analysis, we developed an iterative algorithm for finding $T_{rot}$ and OPR. For each iteration:

1. The rotational temperature $T_{rot}(i)$ is obtained for an assumed OPR($i$) value (the index "$i$" designates the number of iteration)

2. A series of fluorescence models is applied, with the rotational temperature set equal to $T_{rot}(i)$ and the ortho-para ratio varied as a free parameter. The new OPR($i + 1$) value is obtained by minimizing the variance around <F/g($T_{rot}(i)$, OPR)>. F/g($T_{rot}(i)$, OPR) is the ratio of observed line flux and fluorescence g-factor[4] for an individual line, and the mean value (<F/g>) is averaged over the observed $H_2O$ lines.

This iterative process converges to an optimal and unique solution for $T_{rot}$ and OPR. We verified that the final solution is not sensitive to the initially assumed values of both parameters.

<FIGURE 2>

<FIGURE 3a, FIGURE 3b>

4.2 *Spatial Profiles of $H_2O$ Rotational Temperature.*

Figure 2 shows the spatial distribution of $H_2O$ emission intensity along the spectrometer entrance slit. The rotational temperature can be determined reliably if the observed transitions sample quantum states whose internal rotational energies span a sufficiently broad range to constrain the distribution. The spectra in the LA⁻ (especially) and LA⁺' settings satisfy this criterion. We extracted five separate measurements of $T_{rot}$ corresponding to different projected distances from the intensity peak[5], as indicated on Fig. 2.

---

[4] The fluorescence efficiency (*g*) is expressed relative to total $H_2O$ content [W ($H_2O$ molecule)$^{-1}$] and depends on $T_{rot}$ and OPR. The g-factors for "hot-band" emission were described in Dello Russo et al. (2004), and were extended by Dello Russo et al. (2005) using Einstein's A-coefficients from Barber et al. (2006).

The measured $T_{rot}$ distributions in LA⁺' (Figure 3a) and LA⁻ (Figure 3b) show very similar trends. We show stochastic errors for individual temperature measurements, to emphasize the variation with projected distance from the intensity peak. These stochastic uncertainties are small because (1) the signal-to-noise ratios of most sampled lines are high, and (2) many lines are sampled. The two temperature distributions (LA⁺', LA⁻) are also subject to systematic uncertainties (see Dello Russo et al. 2005, 2006; Bonev 2005, Chapter 4; Bonev et al. 2007), estimated to be ± 4 K for each setting.

4.3    *Spatially-resolved Measurements of the H₂O Ortho-Para Ratio.*

Our OPR measurements are shown as diamonds on Figure 3a (LA⁺') and Figure 3b (LA⁻). Stochastic error bars (1σ) are shown for each measurement. For each setting, the weighted mean ortho-para ratio (<OPR>) is indicated as a solid line and its stochastic and systematic uncertainties (discussed in §5.4) are shown as dotted and dashed lines respectively. For LA⁺' we obtain <OPR> = 3.28 ± 0.25, while for LA⁻ we obtain <OPR> = 3.05 ± 0.26 (systematic errors). Assuming that the OPR did not change during the measurement interval, combining LA⁺' and LA⁻ yields a mean weighted value <OPR> = 3.17 ± 0.20. Within the uncertainty, we conclude that the ortho-to-para ratio in SW3-B is consistent with the statistical equilibrium value of 3.0, with $T_{spin} > \sim 45$K. We see no evidence for OPR change in the coma.

---

[5] Emission from a parent volatile commonly peaks at the nucleus.

5. **Discussion of $T_{rot}$ and OPR Results**

The $T_{rot}$ and OPR measurements test the quality of extant $H_2O$ fluorescence models. The bottom panels in Figure 1a, 1b, and 1c show modeled $H_2O$ spectra, for the $LA^{+'}$, $LA^0$, and $LA^-$ setting respectively. These models are calculated for OPR = 3.0 and $T_{rot}$ is taken from the analysis of $LA^{+'}$ and $LA^-$ ($T_{rot}$ = 110 K is adopted for $LA^0$). The overall correlations between modeled and observed spectra are > 95% for the nucleus-centered extracts in $LA^{+'}$ and $LA^-$. Two of the brightest lines sampled in the $LA^0$ setting (at rest frequencies ~3394 and ~3373 cm$^{-1}$) cannot be adequately modeled. The flux to g-factor ratio (F/g) for these lines was also abnormally high in several other comets, suggesting significantly underestimated fluorescence efficiency and/or blends with unknown lines.

Limitations of the fluorescence models were discussed extensively elsewhere (Dello Russo et al. 2004, 2005, 2006; Bonev 2005 (Chapter 4); Bonev et al. 2007). Systematic sources of error in $T_{rot}$ and (especially) the OPR are quantified by the line-by-line scatter in the flux to g-factor ratio (F/g) (see insets in Figures 1a-1c) and might include (slight) inaccuracies in predicted line frequencies, g-factors, and/or corrections for telluric extinction. In addition, while each $T_{rot}$ measurement provides an "effective" value for a given aperture, the actual temperature varies radially from the comet nucleus. Some of the scatter in (line-by-line) F/g values and in the OPR spatially-resolved measurements (Fig. 3a, 3b) might be attributed to this effect, if the relative intensities of H$_2$O lines deviate from those predicted by assuming a unique (for a given aperture) temperature.

5.1  *Temperature Variation in the Inner Coma.*

There are a number of model predictions for the temperature structure of cometary comae (for example see the reviews of Combi et al. 2004, Gombozi et al. 1986, and references therein; Combi et al. 1999; Combi 1989; Ip 1989; Combi and Smyth 1988; Bockelee-Morvan & Crovisier 1987; Marconi and Mendis 1982). Our $T_{rot}$ measurements might be valuable for qualitative[6] comparison with these models, particularly for the innermost ~30 km from the nucleus. In this coma region, the radial temperature gradient is predicted to be only weakly dependent on the gas production rate (Bockelee-Morvan & Crovisier 1987). Gas temperature profiles modeled for heliocentric distances of ~ 1 AU (Combi & Smyth 1988; Combi et al. 1999) predict temperatures of ~ 100 K near the nucleus and a temperature decrease (attributed mainly to near-adiabatic cooling) of several tens of degrees, similar to the observations reported here. Neutral-neutral collisions become less frequent with increasing nucleocentric distance, and electron collisions then become the controlling agent for thermalizing the H$_2$O rotational distribution (10$^3$-10$^4$ km, Xie and Mumma 1992). Depending on both gas production rate and solar UV flux density, the temperature is predicted to increase in outer regions of the coma due to heating from thermalized H-atoms, created by water photolysis. However, the SW3 observations do not sample those outer ($> 10^2$ km) regions.

---

[6] There are two main reasons for the qualitative nature of the comparison. First, the $T_{rot}$ measurements (Fig. 3a, 3b) do not correspond to unique cometocentric distances. Instead, each measurement is an "effective" temperature for the given range of projected distances from the peak gas intensity and for the full range of cometocentric distances sampled in the corresponding gas column along the line of sight. Second, models typically assume only water release directly from the nucleus. We cannot rule out that SW3-B had a secondary source of sublimating H$_2$O molecules in near-nucleus coma, since the comet was especially active, releasing chunks of bulk material from its surface.

5.2    *Ortho-Para Ratio Consistent with Statistical Equilibrium.*

Dello Russo et al. (2007) observed SW3-B with NIRSPEC on Keck 2 and reported OPR = 3.2 ± 0.3, on both UT 14 and 15 May 2006. They did not report spatially-resolved measurements, but their result agrees well with the OPRs reported in this study for individual settings (LA$^{+\prime}$ and LA$^{-}$) and therefore with our weighted mean (3.17 ± 0.20). Previous OPR measurements for $H_2O$ in comets were recently reviewed in Bonev et al. (2007). Notably, the OPR observed in SW3-B agrees with some other comets (e.g., C/2004 Q2 (Machholz), C/1986 P1 (Wilson), but contrasts with results from others that are consistent with lower values (e.g., C/1995 O1 (Hale-Bopp) and 1P/Halley). The spin temperature in the former group exceeds 35-40 K, while in the latter examples $T_{spin}$ is clustered between 25-35 K (see Figure 1 in Bonev et al. 2007).

6.    **Search for Temporal Variations in the $H_2O$ Production Rate**

Paper I reported a rotational temperature (only for a nucleus-centered spectral extract) and a water production rate (1.9 ± 0.2 x $10^{28}$ molecules s$^{-1}$), that were extracted after combining the spectra from all three instrument settings used during the night. Here, we search for temporal variability in Q($H_2O$) during the ~3 hour time frame of our observations. Our methodology for retrieving Q($H_2O$) has been described in detail in DiSanti et al. (2001, 2006), Bonev et al. 2006 and references therein. We measured the $H_2O$ production rates by analyzing separately each individual setting, first extracting a rotational temperature (§4) and then total production rates from the lines measured. For LA$^{+'}$ and LA$^-$ we used the $T_{rot}$ values retrieved for the nucleus-centered extracts (107 K and 110 K, respectively). The transitions in LA$^0$ do not constrain $T_{rot}$ well because of insufficient spread in rotational excitation. For this setting, we adopted the temperature found in the LA$^-$ setting (observed immediately after LA$^0$, see Table 1), and we verified that the result for Q(H$_2$O) measured in LA$^0$ does not change significantly if the temperature found in LA$^{+'}$ is used instead. We also verified that the Q(H$_2$O) measurements in LA$^-$ and LA$^{+'}$ are not affected by varying $T_{rot}$ within ranges that exceed its systematic uncertainty of 4 K.

The production rates extracted from each individual setting are shown in Table 1. The measurement in LA$^0$ and LA$^-$ deviate by less than 1σ from the result reported in Paper I, while the measurement in LA$^{+'}$ is slightly lower, although definitive evidence for variability cannot be argued given the size of the measurement error.

The present results do not support significant variability in water production within the ~3 hour span of our observation, even though the comet was observed after an outburst and the gas productivity was found to decrease by a factor of ~1.8 from 9 May to 10 May 2007 (see Paper I). The lack of strong evidence for temporal variability in Q(H$_2$O) affirms the assumption in Paper I of constant gas productivity, and the mixing ratios of organics with respect to H$_2$O stand as reported by Kobayashi et al. At the same time, we note that although substantially improved from Paper I, the time scale of our measurements (see Table 1) is still significantly larger than the resident time of a molecule within the aperture.

7. **Summary**

Comet 73P-B/Schwassmann-Wachmann 3 was observed with IRCS/Subaru at geocentric distance of 0.074 AU. Multiple $H_2O$ emission lines were detected in non-resonant ("hot-band") fluorescence. The $H_2O$ production rate was measured for three separate time intervals and remained relatively stable during the time span of our observations (Table 1). The $H_2O$ rotational temperature and ortho-para ratio were measured for several positions in the coma. Our principal results are:

1. The $T_{rot}$ spatial distributions extracted from two different time frames show very similar trends. We find a decrease in $T_{rot}$ from ~110 K to ~90 K as the projected distance from the nucleus increased from ~5 to ~30 km.

2. The $H_2O$ ortho-para ratio is consistent with the statistical equilibrium value (3.0) for all spatially resolved measurements (weighted mean OPR = $3.17 \pm 0.20$). This implies a nuclear spin temperature higher than ~ 45 K.

**Acknowledgements**

This paper is based on data collected at Subaru Telescope, which is operated by the National Astronomical Observatory of Japan. The observations can be accessed through the Subaru Telescope Archive System (https://stars.naoj.org). This work was financially supported by the Ministry of Education, Science, and Culture, Grant-in-Aid for Young Scientists 19740107 (H.K.), and by the NASA's Planetary Astronomy and Astrobiology Programs (M.J.M.). The authors wish to recognize and acknowledge the very significant cultural role and reverence that the summit of Mauna Kea has always had within the indigenous Hawaiian community. We are most fortunate to have the opportunity to conduct observations from this mountain.

**References**

Barber, R. J., Tennyson, J., Harris, G. J., & Tolchenov, R. N., 2006. A high-accuracy computed water line list. MNRAS 368, 1087-1094. DOI: 10.1111/j.1365-2966.2006.10184.

Bensch, F. & Bergin, E.A., 2004. The Pure Rotational line Emission of Ortho-Water vapor in Comets: I. Radiative Transfer Model. Astrophys. J. 615, 531-544. Erratum: Astrophys. J. 659, 1795-1799 (2007).

Bockelée-Morvan, D. & Crovisier, J., 1987. The 2.7-micron water band of comet P/Halley - Interpretation of observations by an excitation model. Astron. Astrophys. 187, 425-430.

Bockelée-Morvan D. and Crovisier J. (1987) The role of water in the thermal balance of the coma. In *Proceedings of the Symposium on the Diversity and Similarity of Comets* (E. J. Rolfe and B. Battrick, eds.), pp. 235–240. ESA SP-278, Noordwijk, The Netherlands.

Bonev, B.P., 2005. Towards a chemical taxonomy of comets: Infrared spectroscopic methods for quantitative measurements of cometary water (with an independent chapter on Mars polar science). Ph.D. thesis (Univ. of Toledo). http://astrobiology.gsfc.nasa.gov/Bonev_thesis.pdf.

Bonev, B.P., Mumma, M.J., DiSanti, M.A., Dello Russo, N., Magee-Sauer, K., Ellis, R.S., & Stark, D.P., 2006. A comprehensive study of infrared OH prompt emission in two comets. I. Observations and effective g-factors. Astrophys. J. 653, 774-787.

Bonev, B.P., Mumma, M.J., Villanueva, G.L., DiSanti, M.A., Ellis, R.S., Magee-Sauer,

K., & Dello Russo, N., 2007.  A Search for variation in the $H_2O$ Ortho-Para ratio and Rotational Temperature in the inner coma of Comet C/2004 Q2 (Machholz). Astrophys. J. 661, L97-L100.

Chapovsky, P.L., & Hermans, L.J.F., 1999.  Nuclear Spin Conversion in Polyatomic Molecules. Ann. Rev. Phys. Chem. 50, 315-345.

Combes, M., and 18 co-authors, 1986.  Infrared sounding of comet Halley from Vega 1. Nature 321, 266-268.

Combi, M. R., Cochran, A. L., Cochran, W. D., Lambert, D. L., and Johns-Krull, C. M. 1999.  Observations and Analysis of High-resolution optical line profiles in Comet Hyakutake (C/1996 B2). Astrophys. J. 512, 961-988.

Combi, M. R., Harris, W. M., and Smyth, W. H. 2004.  Gas Dynamics in the Cometary Coma: Theory and Observations.  In Comets II (M. C. Festou et al., eds., University of Arizona Press, Tucson), 523-552.

Combi, M. R. and Smyth, W. H. 1988.  Monte-Carlo Particle-trajectory Models for Neutral Cometary Gases.   I.  Models and Equations. Astrophys. J. 327, 1-26-1043

Combi, M. R. 1989.  The outflow speed of the coma of Halley's Comet.  Icarus 81, 41-50.

Crovisier J., 1984.  The water molecule in comets: Fluorescence mechanisms and thermodynamics of the inner coma.  Astron. Astrophys. 130, 361-372.

Crovisier, J., 2006.  New trends in cometary chemistry.  In: Chemical Evolution of the Universe, Faraday Discussion 133, 375-385.

Crovisier, J., Leech, K., Bockelée-Morvan, D., Brooke, T. Y., Hanner, M. S., Altieri, B., Keller, H. U., & Lellouch, E., 1997.  The spectrum of Comet Hale-Bopp (C/1995 O1) observed with the Infrared Space Observatory at 2.9 AU from the Sun. Science 275, 1904-1907.

Dello Russo, N., DiSanti, M.A., Magee-Sauer, K., Gibb, E.L., Mumma, M.J., Barber, R.J., & Tennyson, J., 2004. Water production and release in comet 153P/Ikeya-Zhang (2002 C1): Accurate rotational temperature retrievals from hot-band lines near 2.9 μm. Icarus 168, 186-200.

Dello Russo, N., Bonev, B.P., DiSanti, M.A., Mumma, M.J., Gibb, E.L., Magee-Sauer, K., Barber, & R.J., Tennyson, J., 2005. Water production rates, rotational temperatures, and spin temperatures in Comets C/1999 H1 (Lee), C/1999 S4, and C/2001 A2. Astrophys. J. 621, 537-544.

Dello Russo, N., Mumma, M. J., DiSanti, M. A., Magee-Sauer, K, Gibb, E. L., Bonev, B. P., McLean, I. S., & Xu, Li-H., 2006. A high-resolution infrared spectral survey of Comet C/1999 H1 (Lee). Icarus 184, 255.

Dello Russo, N., Vervack, R.J., Weaver, H.A., Biver, N., Bockelée-Morvan, D., Crovisier, J., & Lisse, C.M., 2007. Compositional homogeneity in the fragmented comet 73P/Schwassmann-Wachmann 3. Nature 448, 172-175. doi:10.1038/nature05908.

DiSanti, M.A., Mumma, M.J., Dello Russo, N., Magee-Sauer, K., Novak, R., Rettig, T.W., 2001. Carbon monoxide production and excitation in Comet C/1995 O1 (Hale–Bopp): Isolation of native and distributed CO sources. Icarus 153, 361–390.

DiSanti, M.A., Bonev, B.P., Magee-Sauer, K., Dello Russo, N., Mumma, M.J., Reuter, D.C., & Villanueva, G.L., 2006. Detection of formaldehyde emission in Comet

C/2002 T7 (LINEAR) at infrared wavelengths: Line-by-line validation of modeled fluorescent intensities. Astrophys. J. 650, 470–483.

DiSanti, M.A., Anderson, W.M., Villanueva, G.L., Bonev, B.P., Magee-Sauer, K., Gibb, E.L., & Mumma, M.J. 2007. Depleted Carbon Monoxide in the Jupiter-Family Comet 73P/Schwassman-Wachmann 3-C. Astrophys. J. 661, L101-L104.

Edwards, D.P. 1992. GENLN2: A general line-by-line atmospheric transmittance and radiance model, Version 3.0 description and users guide, NCAR/TN-367-STR, National Center for Atmospheric Research, Boulder, CO.

Festou, M.C., Keller, H.U., & Weaver, H.A., 2004. A Brief Conceptual History of Cometary Science. In: Comets II (University of Arizona Press, Tucson), 3-16.

Fuse, T., Yamamoto, N., Kinoshita, D., Furusawa, H., and Watanabe, J.-I., 2007. Observations of Fragments Split from Nucleus B of Comet 73P/Schwassmann–Wachmann 3 with Subaru Telescope. Publ. Astron. Soc. Japan 59, 381–386.

Gombosi, T. I., Nagy, A. F., and Cravens, T. E. 1986. Dust and neutral gas modeling of the inner atmospheres of Comets. Rev. Geophys. 24, No. 3, 667-700.

Ip. W.- H. 1989. Photochemical Heating of Cometary comae. III. The radial variation of the expansion velocity of CN shells in Comet Halley. Astrophys. J. 346, 475-480.

Kawakita, H., Dello Russo, N., Furusho, R., Fuse, T., Watanabe, J.-I., Boice, D.C., Sadakane, K., Arimoto, N., Ohkubo, M., & Ohnishi, T., 2006. Ortho-to-Para Ratios of Water and Ammonia in Comet C/2001 Q4 (NEAT): Comparison of Nuclear Spin Temperatures of Water, Ammonia, and Methane. Astrophys. J. 643, 1337-1344.

Kobayashi, N., Tokunaga, A., Terada, H., Goto, M., Weber, M., Potter, R., Onaka, P., Ching, G., Young, T., Fletcher, K., Neil, D., Robertson, L., Cook, D., Imanishi, M., Warren, D., 2000. IRCS: infrared camera and spectrograph for the Subaru Telescope. SPIE 4008, 1056-1066.

Kobayashi, H., Kawakita, H., Mumma, M.J., Bonev, B.P., Watanabe, J.-I., & Fuse, T. 2007. Organic volatiles in comet 73P-B/Schwassmann-Wachmann 3 observed during its outburst: A clue to the formation region of the Jupiter-family comets. Astrophys. J. Letters (in press, August 2007). Paper I.

Lecacheux, A., and 21 colleagues, 2003. Observations of water in comets with Odin. Astron. Astrophys. 402, L55–L58.

Limbach, H.-H., Buntkowsky, G., Matthes, J., Gründemann, S., Pery, T., Walaszek, B, & Chaudret, B., 2006. Novel Insights into the Mechanism of the Ortho-para Spin Conversion of Hydrogen Pairs: Implications for Catalysis and Interstellar Water. ChemPhysChem. 7, 551-554. DOI: 10.1002/cphc.200500559

Marconi, M. L. and D. A. Mendis 1982. A multi-fluid model of an $H_2O$-dominated dusty cometary atmosphere. The Moon and the Planets, 27, 431-452.

Mumma, M.J., Weaver, H.A., and Larson, H.P., Davis, D.S. & Williams, M., 1986. Detection of Water Vapor in Halley's Comet. Science 232, 1523-1528. ibid 234: 128.

Mumma, M.J., Weaver, H.A., and Larson, H.P. 1987. The Ortho-Para Ratio of Water Vapor in Comet Halley. Astron. Astrophys. 187, 419-424.

Mumma, M.J., Weissman, P. R., & Stern, S. A. 1993. Comets and the origin of the solar system - Reading the Rosetta Stone. In: Protostars and planets III (A93-42937 17-90), 1177-1252.

Mumma, M.J., DiSanti, M.A., Tokunaga, A., & Roettger, E.E., 1995. Ground-based detection of water in Comets Shoemaker-Levy 1992 XIX and 6P/d'Arrest: Probing parent volatiles by hot-band fluorescence. Bull. Amer. Astron. Soc. 27, No. 3: 1144.

Mumma, M. J., DiSanti, M. A., Dello Russo, N., Fomenkova, M., Magee-Sauer, K., Kaminski, C. D., & Xie, D. H. 1996. Detection of Abundant Ethane and Methane, Along with Carbon Monoxide and Water, in Comet C/1996 B2 Hyakutake: Evidence for Interstellar Origin. Science 272, 1310-1314.

Rothman, L. S., and 29 coauthors 2005. The HITRAN 2004 molecular spectroscopic database. J. Quant. Spectr. Rad. Tran. 96 (2), 139-204.

Sekanina, Z., 2005. Comet 73P/Schwassmann-Wachmann: Nucleus Fragmentation, Its Light-Curve Signature, and Close Approach to Earth in 2006. Intl. Comet. Quart. 27, 225-240.

Villanueva, G.L., Bonev, B.P., Mumma, M. J., Magee-Sauer, K., DiSanti, M.A., Salyk, C., Blake, G.A. 2006. The Volatile Composition of the split ecliptic comet 73P/Schwassman-Wachmann 3: A comparison of fragments C and B. Astrophys. J. 650, L87-L90.

Weaver, H.A., & Mumma, M.J.,1984. Infrared Molecular Emissions from Comets. Astrophys. J. 276, 782-797.

Weaver, H. A., Mumma, M. J., & Larson, H. P., 1987. Infrared Investigation of Water in comet P/Halley. Astron. Astrophys. 187, 411-418.

Weaver, H.A., Lisse, C.M., Mutchler, M.J., Lamy, P., Toth, I., Reach, W.T., 2006. Hubble Space Telescope Investigation of the Disintegration of 73P/Schwassmann-Wachmann 3. Bull. Amer. Astron. Soc. 38, 490.

Xie, X, & Mumma, M. J. 1992. The effect of electron collisions on rotational populations of cometary water. Astrophys. J. 386, 720-728.

Xie, X, & Mumma, M. J. 1996. Monte Carlo Simulation of Cometary Atmospheres: Application to Comet P/Halley at the Time of the Giotto Spacecraft Encounter. I. Isotropic Model. Astrophys. J. 464, 442-456.

**Table 1**

Observing Log – 73P-B/Schwassmann-Wachmann 3 – UT 2006 May 10

| | $R_h$ = 1.027 AU, $\Delta$ = 0.074 AU [a] | | | | |
|---|---|---|---|---|---|
| Setting | Frequency Range | UT Time [b] | $T_{int}$ | $\Delta$-dot [a] | Q(H$_2$O) [c] |
| | (cm$^{-1}$) | (hh:mm) | (min) | (km s$^{-1}$) | (10$^{28}$ s$^{-1}$) |
| LA$^{+'}$ | 3520 – 3450 | 11:19 – 12:11 | 32 | -6.44 | 1.5 ± 0.1 |
| LA$^{0}$ | 3408 – 3330 | 12:29 – 13:04 | 24 | -6.30 | 1.8 ± 0.2 |
| LA$^{-}$ | 3500 – 3410 | 13:11 – 13.52 | 28 | -6.20 | 1.7 ± 0.2 |

[a] $R_h$, $\Delta$, and $\Delta$-dot are respectively heliocentric and geocentric distance, and radial velocity with respect to the observing site (topocentric).

[b] Universal Time for the beginning and for the end of the observing sequence: includes total integration time on source ($T_{int}$), "overhead" for nodding the telescope (see §2 of Paper I), and reading out and saving the acquired data.

[c] H$_2$O production rate (see §6)

**Figure Captions**

FIG. 1. – $H_2O$ non-resonant fluorescent emission from the inner coma of comet 73P-B/Schwassmann-Wachmann 3. Three instrument settings were used: $LA^{+'}$ (a), $LA^-$ (b), and $LA^0$ (c), as described in Table 1. The $H_2O$ data in each setting are represented by a three-panel graphic. The top panel shows the measured cometary spectrum and the best-fit terrestrial atmospheric transmittance model (dashed line) convolved to the instrumental resolution and normalized to the mean continuum intensity of the comet data. The middle panel shows the residual spectra after removing the telluric absorption. The dashed lines outline the 1σ photon noise envelope. The bottom panel shows modeled $H_2O$ spectra. Transitions of ortho-$H_2O$ and of para-$H_2O$ are shown with red and blue respectively. The frequency ($cm^{-1}$) axis corresponds to the cometary rest frame. Insets: Line-by-line ratios of observed fluxes and modeled g-factors (*F/g*), normalized to their weighted mean value (*<F/g>*). The spread around the weighted mean exceeds the stochastic errors of individual points and is (pending improved models) the limiting source of systematic uncertainties in the $T_{rot}$ and OPR retrievals. The quantity on the x-axis is the average energy of the rotational levels in the ground vibrational state, weighted by their pumping contribution to the population of the upper ro-vibrational level for the corresponding transition (see Dello Russo et al. 2005, and Bonev 2005 [§4.8]).

FIG. 2. – Spatial profile of $H_2O$ emission. Each temperature and OPR measurement (Figs. 3 and 4) corresponds to a range of projected (on the sky plane) distances from the nucleus, as indicated on this graphic. The slit was oriented –45º West of North on the sky with the projected sunward direction ~35º East of North. This spatial profile was constructed by summing (column by column) the signal of all detected water lines in the LA$^{+'}$ setting, thereby forming a spatial profile with high signal-to-noise ratio.

FIG. 3. – Rotational temperatures (stars on Fig. 3a, squares on Fig. 3b) and OPRs (diamonds) measured for $H_2O$ in the LA$^{+'}$ (3a) and LA$^-$ (3b) setting respectively. Individual stochastic errors are shown for each temperature measurement. The systematic uncertainty in the $T_{rot}$ distribution is shown in the lower left corner of each figure. The weighted mean OPR, its stochastic error, and its systematic uncertainty (this uncertainty is expected to affect all points in the same direction) are shown with solid line, dotted lines, and dashed lines respectively.

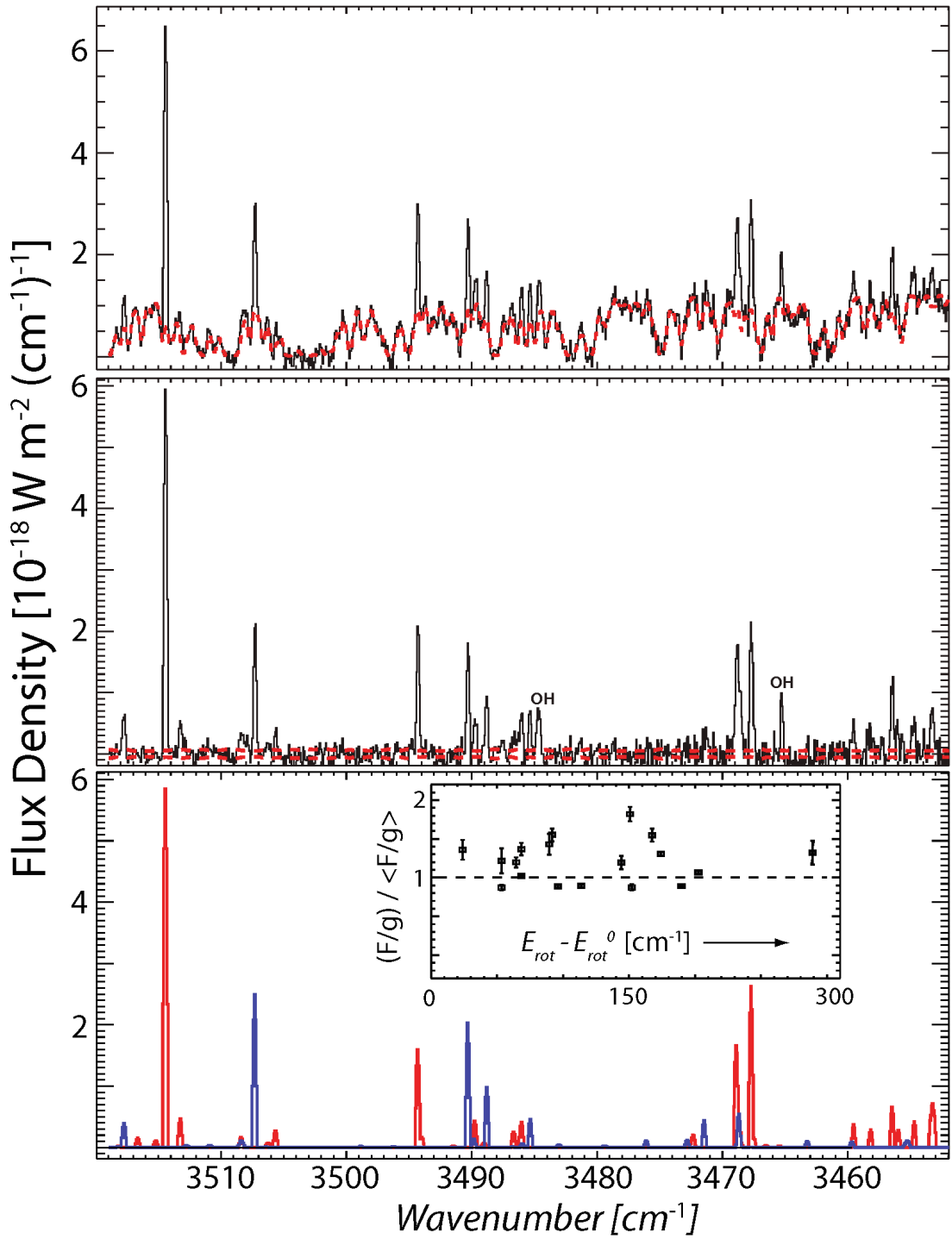

Figure 1a

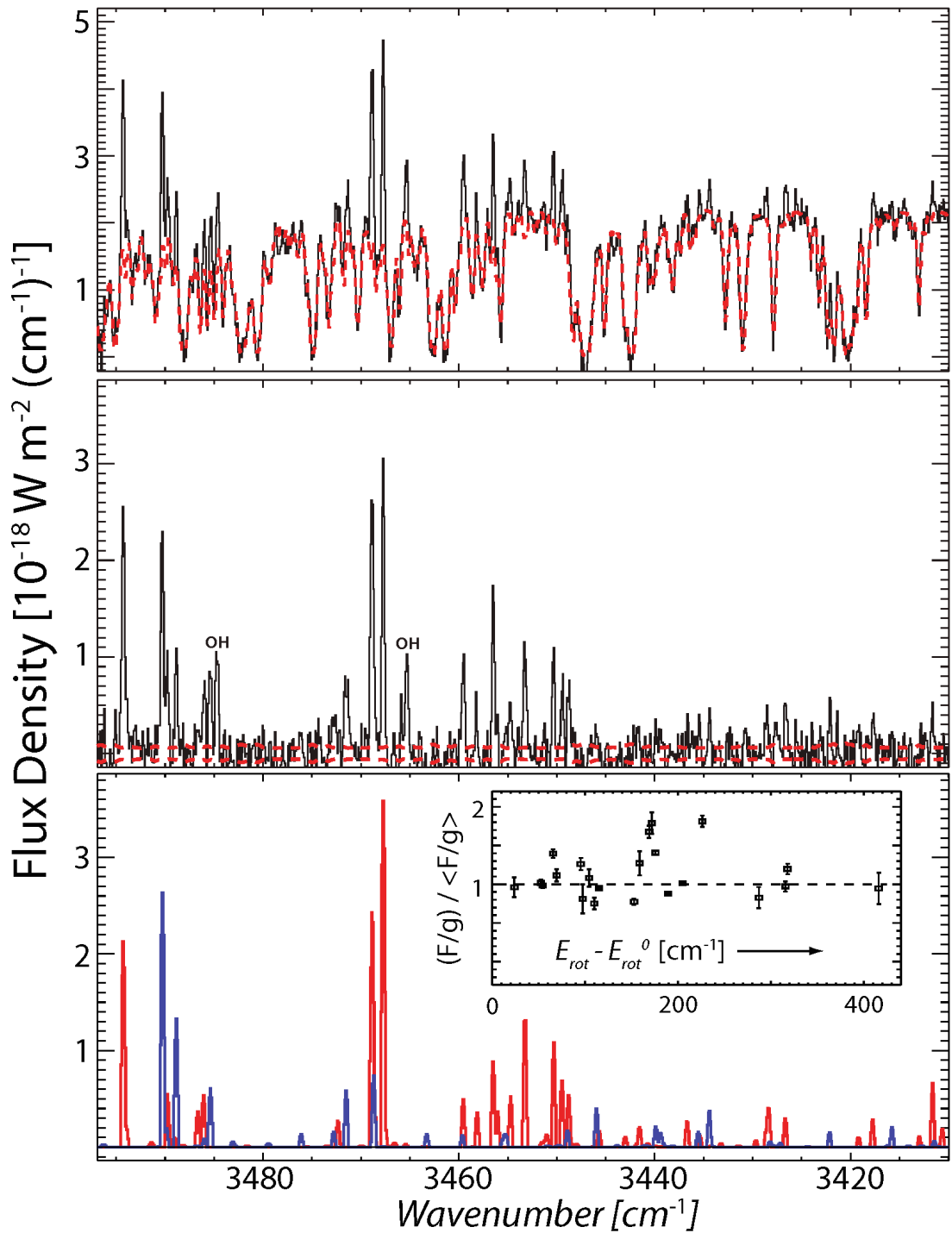

Figure 1b

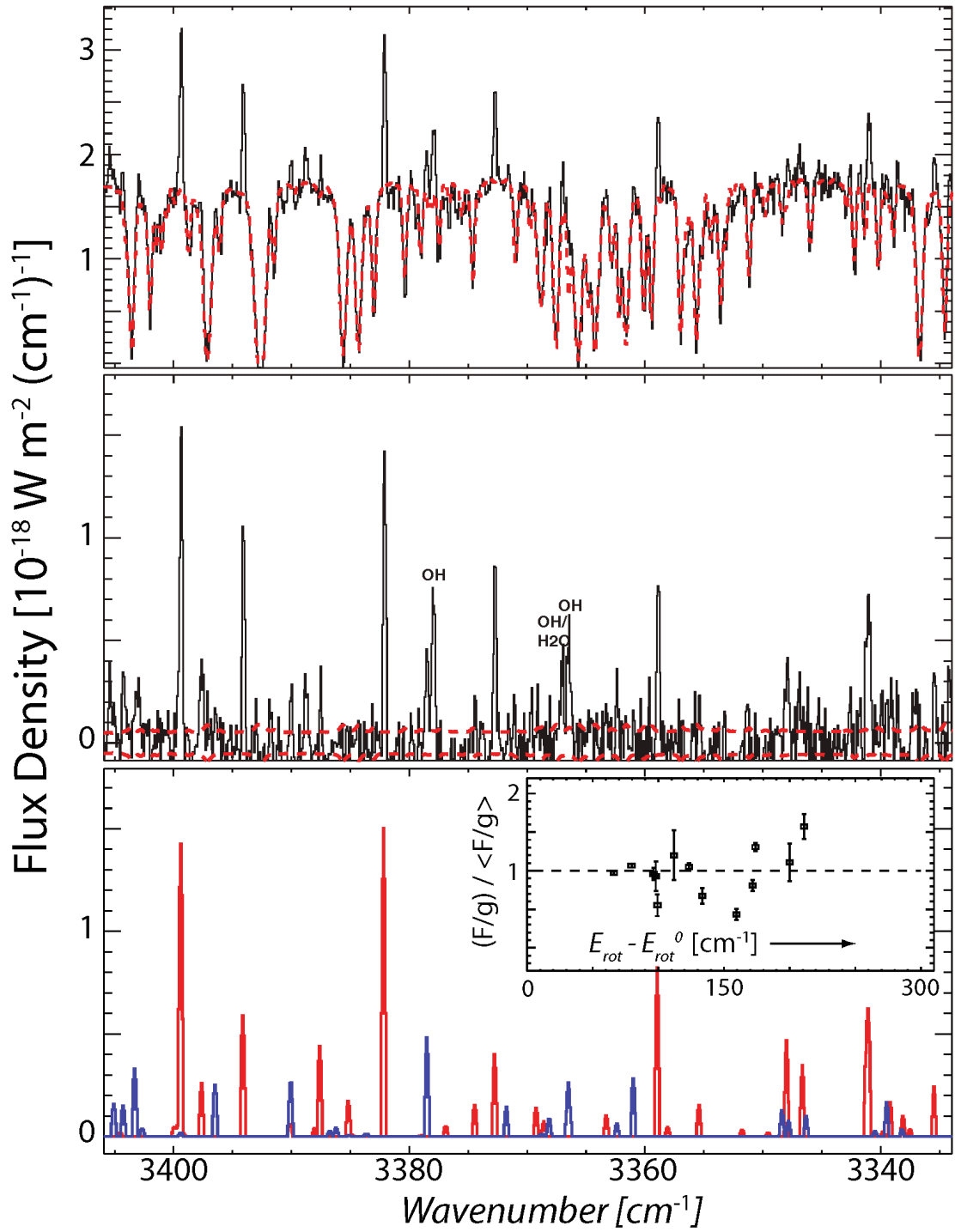

Figure 1c

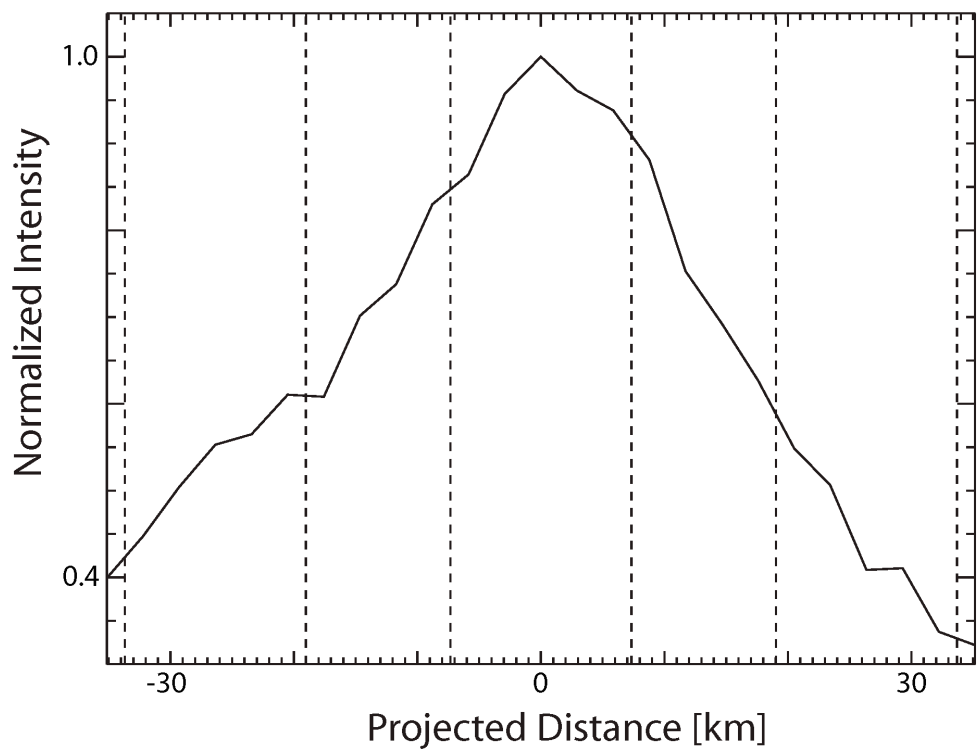

Figure 2

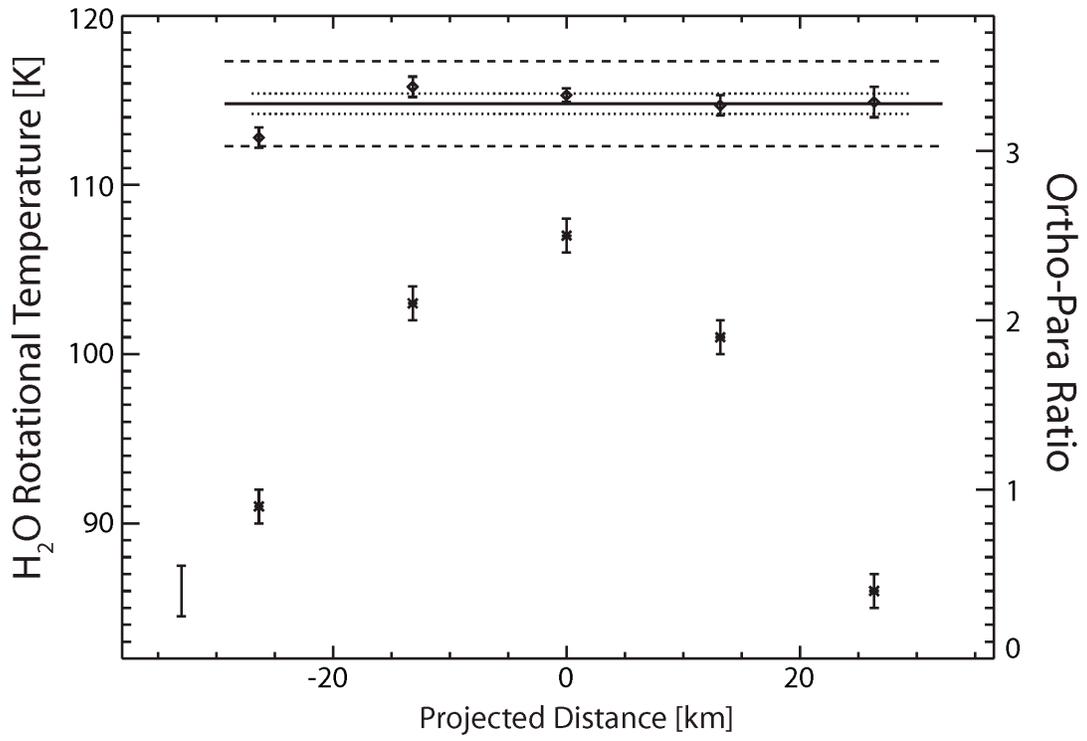

Figure 3a

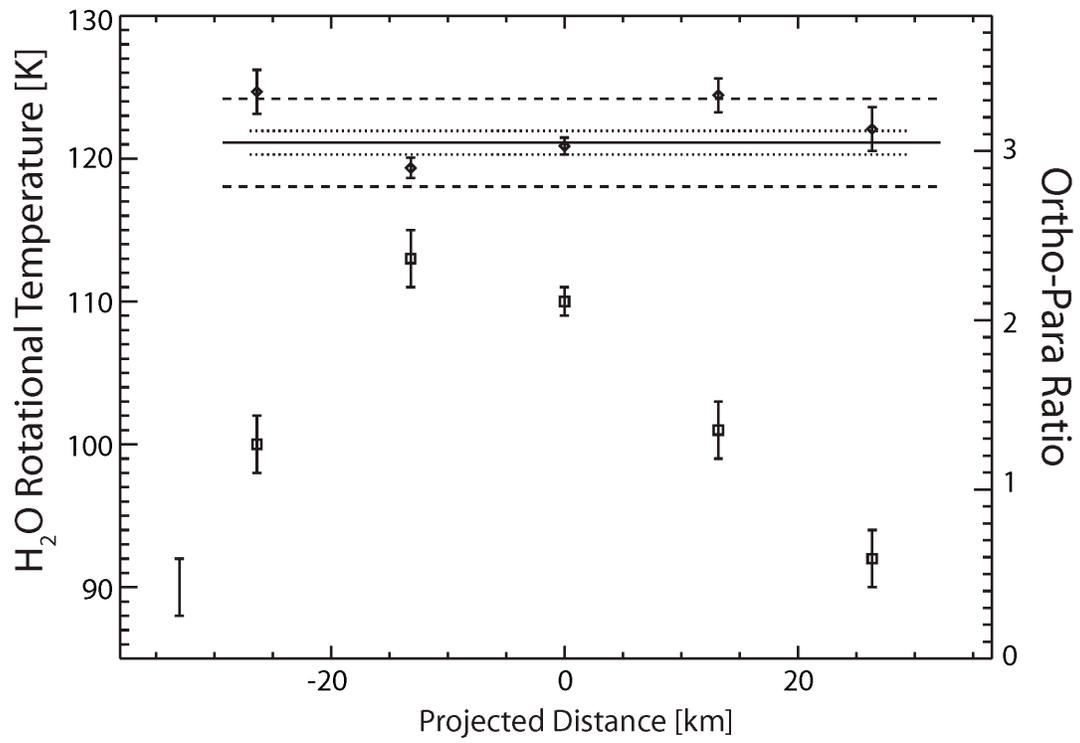

Figure 3b